\documentclass[12pt, letterpaper]{article} 

\usepackage[includeheadfoot,
            marginratio={1:1,2:3}, 
            width=440pt, 
            height=688pt,]{geometry}

\usepackage{amsmath}
\usepackage{amsfonts}
\usepackage{amssymb}
\usepackage{bbm,bm}
\usepackage{mathtools}
\usepackage{slashed}
\usepackage{mathalfa}

\usepackage{graphicx,caption,subfigure}
\usepackage{tikz} 
\usepackage{tikz-cd}
\usetikzlibrary{decorations.pathreplacing,calc,tikzmark}
\usepackage{booktabs}
\usepackage{adjustbox}
\usepackage[utf8]{inputenc}
\usepackage[splitrule]{footmisc} 
 
\usepackage{placeins}
\usepackage{empheq}
\usepackage{paralist}
\usepackage{cite}
\usepackage[normalem]{ulem}
\usepackage{soul}

\usepackage[colorlinks=false,urlbordercolor=red
]{hyperref}
\usepackage{xcolor}
\usepackage{tensor}

\newcommand{\nc}{\newcommand}
\nc{\lb}{\llbracket}
\nc{\rb}{\rrbracket}
\nc{\gl}{\llbracket}
\nc{\gr}{\rrbracket}
\nc{\del}{\partial}
\nc{\eq}[1]{\begin{equation}
                     \begin{split} #1 \end{split}
                     \end{equation}
}
\nc{\ov}{\overline}
\nc{\fa}{\hat}
\nc{\fb}{\MakeUppercase}
\nc{\fc}{\tilde}
\nc{\myhash}{\raisebox{\depth}{\#}}

\allowdisplaybreaks[2]
\numberwithin{equation}{section}
\DeclareUnicodeCharacter{2212}{-}

\begin{document}


\vspace*{-1.5cm}
\begin{flushright}
  {\small
  MPP-2023-192
  }
\end{flushright}

\vspace{1.0cm}
\begin{center}
  {\Large Demystifying the Emergence Proposal } 
\vspace{0.4cm}

\end{center}

\vspace{0.25cm}
\begin{center}
{
    Ralph Blumenhagen$^{1,2}$, Niccol\`o Cribiori$^1$, Aleksandar Gligovic$^{1,2}$, \\[.2cm] Antonia Paraskevopoulou$^{1,3}$
}
\end{center}

\vspace{0.0cm}
\begin{center} 
\emph{
$^{1}$ 
Max-Planck-Institut f\"ur Physik (Werner-Heisenberg-Institut), \\ 
F\"ohringer Ring 6,  80805 M\"unchen, Germany } 
\\[0.1cm] 
\vspace{0.25cm} 
\emph{$^{2}$ Exzellenzcluster ORIGINS, Boltzmannstr. 2, D-85748 Garching, Germany}\\[0.1cm]
\vspace{0.25cm} 
\emph{$^{3}$ Ludwig-Maximilians-Universit{\"a}t M\"unchen, Fakult{\"a}t f{\"u}r Physik,\\ 
Theresienstr.~37, 80333 M\"unchen, Germany}\\[0.1cm]
\vspace{0.3cm}
\end{center} 
\vspace{0.5cm}

\begin{abstract}
We revisit the Emergence Proposal in the vector multiplet moduli space of type IIA $N=2$ supersymmetric string vacua in four dimensions, for which the string tree-level prepotential and the string one-loop correction are exactly known via mirror symmetry. 
We argue that there exists an exact notion of emergence, according to which these four-dimensional couplings can be computed exactly in any asymptotic limit in field space.
In such limits, a perturbative quantum gravity theory emerges, whose fundamental degrees of freedom include all complete infinite towers of states with typical mass scale not larger than the species scale. 
For a decompactification limit, this picture is closely related to and in fact motivated by the computation of Gopakumar-Vafa invariants. 
In addition, in the same limit our results suggest that the emergent theory will also contain asymptotically tensionless wrapped $NS5$-branes.
\end{abstract}

\thispagestyle{empty}
\clearpage

\setcounter{tocdepth}{2}



\section{Introduction}

Among the various swampland conjectures that have been proposed in the endeavor of revealing the underlying principles of quantum gravity, see \cite{Palti:2019pca,vanBeest:2021lhn,Grana:2021zvf,Agmon:2022thq} for reviews, the so-called Emergence Proposal \cite{Heidenreich:2017sim,Grimm:2018ohb,Heidenreich:2018kpg,Corvilain:2018lgw} is probably the least established and understood one. 
It is closely tied to the swampland distance conjecture \cite{Ooguri:2006in} and suggests that terms in the low-energy effective action emerge at one-loop from integrating out towers of states becoming exponentially light in asymptotic regions of the moduli space.
When following an effective field theory (EFT) approach\footnote{We thank the authors of \cite{Grimm:2018ohb} for reminding us that this EFT approach was only employed as an intermediate tool until an ultimate quantum gravity approach will be available.}, one usually integrates out those states whose masses are below the species scale \cite{Dvali:2007hz,Dvali:2007wp}, which is considered as the ultraviolet cut-off of quantum gravity. 

For decompactification limits, or limits dual to them, it was shown in some examples \cite{Marchesano:2022axe,Castellano:2022bvr} that the leading order singular behavior of field metrics and gauge couplings comes out qualitatively correct by such a computation. 
However, a couple of puzzling issues have been reported recently in \cite{Blumenhagen:2023yws}, in a set-up involving more than one modulus and especially in so-called emergent string limits \cite{Lee:2019wij}, where the lightest tower of states includes also a string becoming tensionless asymptotically (we will refer to it as a string tower). 
Here, one encounters first of all an issue with the very definition of the species scale, since, for emergent string limits, quantum field theory computations lead to results different from those obtained with black hole arguments. 
In the first case, one gets a logarithmic enhancement factor, whose role is unclear. 
Besides, even if qualitatively part of the structure of the four-dimensional effective action does emerge, in the fine print there are a couple of shortcomings, such as: incorrect (relative) numerical coefficients, extra logarithmic factors for emergent string limits, and additional terms at subleading order with no clear interpretation. 
We consider this as evidence that the precise formulation of the Emergence Proposal in quantum gravity is still unclear.

In this paper, we take a critical look at the Emergence Proposal and, in the course of our analysis, we develop a general picture leading to exact emergence of all terms in the effective action.
From this perspective, previous results using the field theory approach to the Emergence Proposal constitute only an incomplete part of this more sophisticated computation which leads to exact emergence, going beyond field theoretic reasoning and involving the full quantum theory of gravity. 
This is also close in spirit to the recent discussion in \cite{Anchordoqui:2023laz}.

Crucially, this step is achieved by summing over all infinite states in the towers and not only up to those states whose masses are below the species scale. While this intuition is behind the proposal in \cite{Palti:2019pca,Grimm:2018ohb}, it has not been pursued quantitatively so far.
The towers to be considered are those with a \emph{typical mass scale} parametrically not larger than the species scale.
In a weakly coupled regime where also the quantization scheme of the theory is known, such typical mass scale is simply the  mass gap between consecutive levels of the tower. 
At strong (string) coupling, where the quantization is usually not known and the former weakly coupled states might migrate across the mass spectrum or even become unstable, we still assume that the naive mass scale, given for instance by the tension of a string, is a good measure for the typical mass scale of the tower. 
This is certainly true for BPS states, whose weakly coupled mass
can be trusted also for strong coupling. 
The situation becomes clearer in a weakly coupled dual description, whenever it exists.

For concreteness, we work in a moduli space with eight preserved supercharges whose asymptotic structure is well-understood, namely the vector multiplet moduli space of type IIA string theory on a Calabi-Yau threefold.  
Infinite distance limits in this setup have been studied and classified in \cite{Corvilain:2018lgw, Lee:2019wij}. In this paper, we will consider two kinds of such limits, namely decompactification and emergent string limits, which are believed to be the only two possibilities according to the emergent string conjecture \cite{Lee:2019wij}.
Moreover, via mirror symmetry the exact form of a large portion of the $N=2$ low-energy effective action is known. Our focus here will be on the string tree-level prepotential $\mathcal F_0$ and the string one-loop correction $\mathcal F_1$, which are closely related to amplitudes of the associated topological string.

Our strategy is the following. 
We consider an asymptotic limit in the  K\"ahler moduli space, in which a certain K\"ahler modulus $t$, or more concretely its   vacuum expectation value $t_0=\langle t \rangle$, is sent to infinity. 
This limit is taken such that the four-dimensional dilaton residing in a hyper multiplet stays fixed.
This guarantees that the four-dimensional Planck-scale does not diverge, so that we can meaningfully talk about (quantum) gravity in four dimensions. 
We propose that in each of these limits a perturbative quantum gravity theory emerges with a characteristic set of fundamental degrees of freedom.
They consist of the complete and infinite towers of states whose
typical mass scale is not larger than the species scale, $\Delta m \leq \tilde \Lambda$. 
Note that it is not the mass of each state itself that is required to be not larger than the species scale, but the typical mass scale  of each tower, which then contributes infinitely many individual states.  This prescription is in accordance with \cite{Palti:2019pca} (see footnote $^{46}$),\cite{Montero:2022prj} and with the more recent work \cite{Burgess:2023pnk}.
In addition, we propose that this emergent theory admits a perturbative-like  expansion in the  naturally small parameter $g_E\simeq t_0^{-1} \simeq N_{\rm sp}^{-1}$, where $N_{\rm sp}$ is the number of species related to the species scale as $\tilde \Lambda = M_{\rm pl}/\sqrt{N_{\rm sp}}$, in four dimensions.
By this we mean that similar to standard string theory with $g_s=\langle \exp(\phi)\rangle$, the expansion in the coupling constant of the emergent theory $g_E$ will have perturbative polynomial corrections as  well as non-perturbative exponentially suppressed corrections of the type $\exp(-1/g_E)$.
In general we do not know how to quantize this emergent theory from first principles but, as we will see, in certain limits one can understand it via an existing dual fundamental string theory.

This work is organized as follows. In section \ref{sec:review}, we review the main ideas behind the Emergence Proposal and we recall some properties of the vector multiplet moduli space of type IIA string theory compactified on Calabi-Yau threefolds.

In section \ref{sec:decomp}, we consider a decompactification limit with  the four-dimensional dilaton remaining constant. This is the M-theory limit where the eleventh direction decompactifies with the Calabi-Yau volume staying finite (in M-theory units).
In this case, while $D0$-branes are certainly the lightest degrees of freedom, one finds that towers of BPS bound states of $D0$-branes, wrapped $D2$-branes and wrapped $NS5$-branes, carrying Kaluza-Klein momentum along the Calabi-Yau threefold have a typical mass scale  not larger than the species scale. 
We do not know how to quantize this theory in general, but luckily, thanks to the seminal work of Gopakumar-Vafa \cite{Gopakumar:1998ii,Gopakumar:1998jq}, certain topological quantities can be obtained via Schwinger one-loop integrals.\footnote{Very recently, Gopakumar-Vafa invariants were also considered in the context of the emergent string conjecture in \cite{Rudelius:2023odg}.}
Hence, we can setup an emergence computation of the prepotential and the genus-one free energy.
As in \cite{Gopakumar:1998ii}, this is performed for a single $D2$-brane wrapping an $S^2$ bound to an arbitrary number of $D0$-branes running in the loop. 
This can also be seen as the complete result  for the (non-compact) resolved conifold.\footnote{Whether one can simply sum over all such contributions to get the full prepotential for compact Calabi-Yau manifolds remains to be seen.}
By summing over all the full infinite towers of such states, we are able to recover the exact type IIA expressions, at both genus zero and one, if we consistently treat the (divergent) infinite sums via  $\zeta$-function regularization.
This constitutes a first fully fledged realization of the (strong) Emergence Proposal. We show how in an EFT approach, cutting-off the sums at some finite level can only recover the leading order behavior, but it crucially misses the correct normalization and subleading corrections.

In section \ref{sec:emstringlim}, we comment on an emergent string limit, which requires a $K3$ fibration over a base $\mathbb P^1$, whose volume is sent to infinity.
The light towers include an emergent four-dimensional string arising from an $NS5$-brane wrapped on the $K3$ fiber and a couple of emergent particle-like states arsing from wrapped type IIA $D$-branes.
We do not know how to quantize this four-dimensional string theory, but we can still understand it in terms of its dual, namely the heterotic string on $K3\times T^2$. 
On the quantitative level, since this emergent theory is just another perturbative description of the same theory, the tree-level type IIA prepotential must also be exactly realized in a perturbative expansion in $g\simeq t_0^{-1}\simeq g_H$.
All this is indeed the familiar story of a string-string duality, where certainly no cut on the tower of states is imposed and the ultraviolet cut-off is just the string scale. We realize that in this case the
Schwinger integral gives (only)  the  one-loop correction of the heterotic dual.
This means that the full prepotential is \emph{not emerging} (in the original sense of \cite{Palti:2019pca,Grimm:2018ohb}, which includes the tree level term) via integrating out towers of states up to the species (i.e. string) scale. Whether an increase of the latter ultraviolet cut-off cures this problem of the Emergence Proposal remains to be seen.\footnote{We thank E.~Palti for pointing out this idea to us.}
Here, we are not following this approach but, more conservatively, we always consider the species scale as the quantum gravity cut-off. Our notion of \emph{exact emergence} simply means that all terms in the prepotential will arise via  the correct quantization of the perturbative quantum gravity theory that emerges  in the considered asymptotic field limit. All in all, our analysis shows that \emph{strong emergence}, to be recalled in the next section, can be realized via \emph{exact emergence} but it does not necessarily have to.
In section \ref{sec:conclusion}, we draw our conclusions and point out future directions of research.

\begin{figure}[h]
\centering
\includegraphics[width=0.9\textwidth]{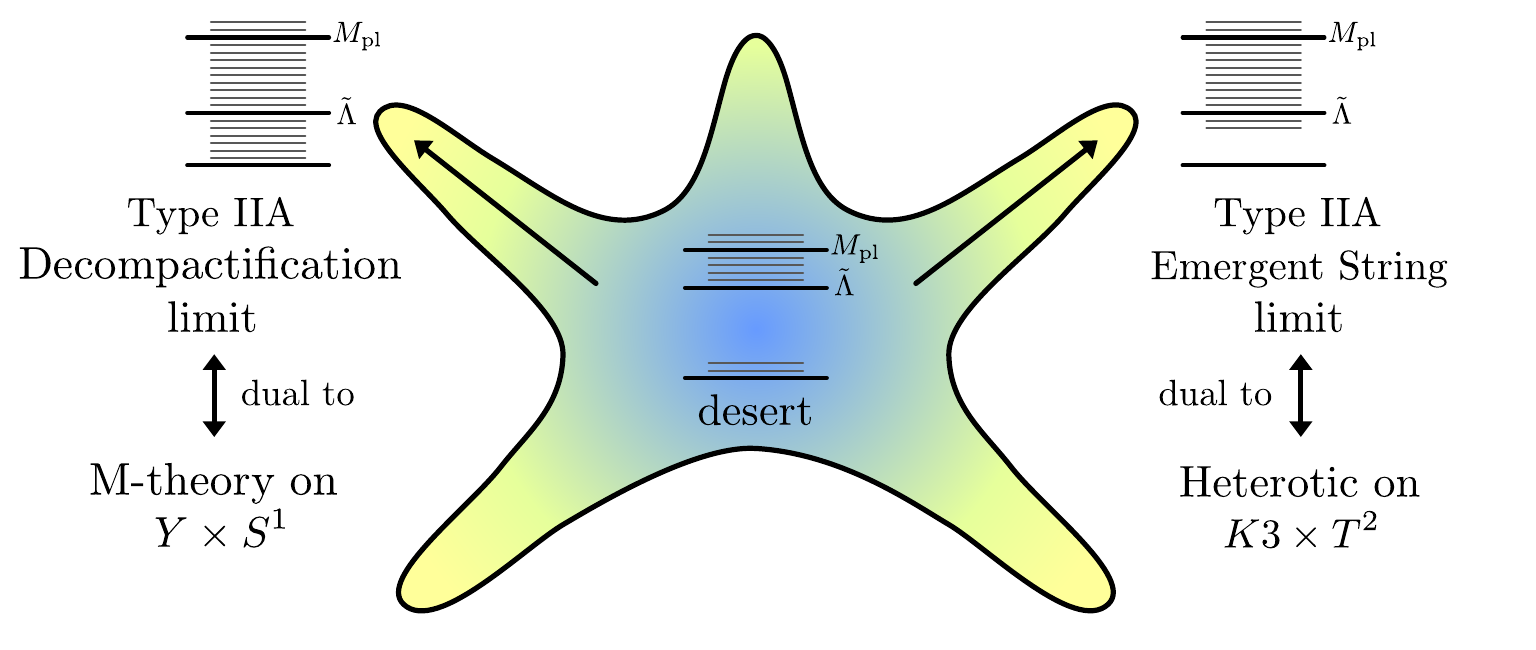}
\caption{Cartoon of type IIA vector multiplet moduli space and asymptotic limits.}
\label{fig:modulispace}
\end{figure}

In summary, as depicted in figure \ref{fig:modulispace}, the type IIA vector multiplet moduli space in four dimensions features a pattern very similar to the familiar duality star in ten/eleven dimensions.
In the figure, we picture the difference between the mass spectrum in the asymptotic limits and the mass spectrum in the bulk region, which has been called the desert in \cite{Long:2021jlv}. 
For the latter, there are only a few light states, while the majority of the towers and the species scale are all close to the Planck scale.
The decompactification limit sheds some new light on a perturbative description of the still mysterious eleven dimensional M-theory, which will be reported in \cite{Blumenhagen:2023xmk}.


\section{Preliminaries}
\label{sec:review}

In this section, we provide a brief review of the original Emergence Proposal and of some salient features of the $N=2$ vector multiplet moduli space of type IIA string compactifications on Calabi-Yau threefolds.

\subsection{The Emergence Proposal}

In view of the swampland distance conjecture\cite{Ooguri:2006in}, which predicts infinite towers of states becoming exponentially light at asymptotic distances in field space, the Emergence Proposal \cite{Heidenreich:2017sim,Grimm:2018ohb,Heidenreich:2018kpg, Corvilain:2018lgw} offers a field-theoretic approach to the effects that such towers have on certain physical quantities. 
In fact, at present two possible ways to interpret the concept of emergence in the context of quantum gravity have been put forward. According to \cite{Castellano:2022bvr}, the strong version of the Emergence Proposal \cite{Palti:2019pca} states the following.
\begin{itemize}
\item  \textbf{Emergence Proposal (Strong)}: In a theory of quantum gravity all light particles in a perturbative regime have no kinetic terms in the ultraviolet. The required kinetic terms appear as an infrared effect due to loop corrections involving the sum over a tower of massless states.
\end{itemize}

\noindent
Clearly, this is a very non-trivial statement which goes beyond our usual understanding of string theory. 
As formulated above, it is not clear if only the kinetic terms of the moduli approaching asymptotic regions are supposed to emerge, or if the full field metric should be recovered, as claimed in \cite{Palti:2019pca}.

There exists a milder formulation, called the weak version of the Emergence Proposal \cite{Castellano:2022bvr}.
\begin{itemize}
\item  \textbf{Emergence Proposal (Weak)}: In a consistent effective theory of quantum gravity, for any singularity at infinite distance in the moduli space, there is an associated infinite tower of states becoming light and inducing quantum corrections to the metrics which match the ‘tree level’ singular behavior.
\end{itemize}

\noindent
This statement is more specific and only refers to the components of the field metric which diverge in the asymptotic limit. In \cite{Blumenhagen:2023yws}, the possibility to extract information for the non-singular components was also explored.

The towers mentioned in both versions of the Emergence Proposal are typically infinite, as predicted by the swampland distance conjecture. 
However, the states that are usually integrated out in an EFT setup are only those lying below the ultraviolet quantum gravity cut-off.  Naively, one would expect the latter to be the four-dimensional Planck mass $M_{\rm pl}$, but it has been proposed \cite{Dvali:2007hz,Dvali:2007wp} (see also \cite{Veneziano:2001ah} for earlier work) that the gravitational cut-off should be lowered down to the species scale $\tilde{\Lambda}$. In four dimensions, this is given by
\begin{equation}\label{species scale def}
\tilde{\Lambda}=\frac{M_{\rm pl}}{\sqrt{N_{\rm sp}}}\,.
\end{equation}
The interpretation of $N_{\rm sp}$ may change if one derives \eqref{species scale def} by using a quantum field theory calculation or rather black hole arguments. In the first case, the species scale is the energy at which the one-loop correction to the graviton propagator, due to the coupling to $N_{\rm sp}$ light states, becomes of the same order as the tree level term. 
Hence, $N_{\rm sp}$ is the number of light states with masses lying below the cut-off $\tilde\Lambda$. 
In the second case, $\tilde\Lambda$ is the inverse of the radius of the smallest possible black hole which can be described by the effective theory. 
Hence, $N_{\rm sp}$ is the entropy of such a black hole. 

A particular feature, which will be relevant to us, is that in the setup we are interested in $N_{\rm sp}$ has been related to the genus-one free energy of the topological string. This has been proposed in \cite{vandeHeisteeg:2022btw} and investigated further in \cite{Cribiori:2022nke,vandeHeisteeg:2023ubh,vandeHeisteeg:2023uxj,Cribiori:2023sch}. Besides, the interpretation of $N_{\rm sp}$ as entropy has been developed into a more complete thermodynamic picture in \cite{Cribiori:2023ffn}.

While in principle we have two ways at our disposal to estimate the species scale, the Emergence Proposal in its standard formulation can be compatible only with a quantum field theory setup. 
Recently, the application of the Emergence Proposal to moduli space metrics and gauge kinetic functions has been studied in detail in \cite{Marchesano:2022axe,Castellano:2022bvr,Castellano:2023qhp}. From these investigations, one learns that e.g.~in four dimensions a gauge coupling $g$ emerges at one loop from integrating out a tower of charged states with mass $m_n=n\Delta m$ and  charge $q_n=n q$, namely
\begin{equation}
\label{gaugeloop}
\frac{1}{g^2}\sim \sum_{n=1}^{N_{\rm sp}}  n^2 q^2 \log\left( {\frac{n\, \Delta m}{\tilde\Lambda}}\right)\, .
\end{equation}
Notice that in an EFT approach the sum stops at $n=N_{\rm sp}$, when indeed the mass of the state is $N_{\rm sp} \Delta m = \tilde \Lambda$. States in the tower with mass larger than $\tilde \Lambda$ are thus not included.
These loop corrections have been evaluated for a couple of asymptotic limits, confirming the weak version of the Emergence Proposal.

A step further has been performed in the recent work \cite{Blumenhagen:2023yws}, where the Emergence Proposal has been tested in an emergent string limit within a toroidal orbifold model with multiple moduli. 
This allowed to compute also the one-loop corrections for the non-asymptotic field directions and to compare them to their known tree-level values.
As a result, it has been pointed out that the quantum field theory approach to the species scale seems to be always plagued by multiplicative logarithmic corrections when string towers are included.
This is inconsistent with the fact that the genus-one topological string free energy receives additive logarithmic corrections.
Moreover, while the one-loop diagrams reproduce the leading order functional form of the gauge couplings and of the moduli metric correctly, the prefactors do not match and also the subleading terms are not related to known string amplitudes.
All of these issues suggest that a complete and precise understanding of the Emergence Proposal is still missing in the literature. 
This is the main motivation behind our investigation.

\subsection{K\"ahler moduli space in type IIA and limits}

The setup we work in is the K\"ahler moduli space of type IIA string theory compactified on a Calabi-Yau threefold. 
We review some of its properties following mainly \cite{Louis:1996ya,Curio:1997si}.

The moduli space of type II string theory compactified on a Calabi-Yau threefold is locally a product of two factors,
\begin{equation}
\mathcal{M}_V \times \mathcal{M}_H\,,
\end{equation}
which do not mix due to supersymmetry. The space $\mathcal{M}_V$ is special K\"ahler and parametri\-zed locally by scalar fields in vector multiplets. The space $\mathcal{M}_H$ is quaternionic and parametri\-zed locally by scalar fields in hypermultiplets. 
We concentrate only on the manifold $\mathcal{M}_V$ associated to type IIA compactifications, for which scalar fields in vector multiplets are complexified K\"ahler moduli. We denote them as $T^i = t^i + i b^i$, with $i=1,\dots, h_{11}(\text{CY}_3)$, where $b^i$ are Kalb-Ramond axions, while $t^i$ are real K\"ahler moduli, defining the K\"ahler cone $t^i>0$. The vector fields are given by the RR $C_3$-form dimensionally reduced on $h_{11}(\text{CY}_3)$ homologically two-cycles. There is one more vector field, the graviphoton, residing in the $N=2$ supergravity multiplet. This is simply given by the type IIA RR $C_1$-form.

The kinetic terms for the K\"ahler moduli and the gauge couplings  are determined by a holomorphic, homogeneous of degree two prepotential $F=F(X)$. 
It is customary to introduce symplectic coordinates $X^\Lambda = (X^0, X^i)$, such that local coordinates on $\mathcal{M}_V$ are recovered as $T^i = -i X^i/X^0$. In our conventions, the K\"ahler potential is then given by $K=-\log\left[ i\left( \ov{X}^\Lambda F_\Lambda-X^\Lambda \ov{F}_\Lambda\right)\right]$.   
Apart from being constrained by supersymmetry, the form of $F(X)$ in type II Calabi-Yau compactifications is restricted by discrete Peccei-Quinn symmetries, which act on $T^i$ and are almost exact for $t^i \gg 1$, and also by non-renormalization theorems. 
Since in these vacua the four-dimensional dilaton resides in a hypermultiplet, the prepotential $F(X)$ is exact at perturbative string tree level and loop corrections can only affect $\mathcal{M}_H$. Nevertheless, there are non-perturbative corrections from worldsheet instantons, such that one can write schematically 
\begin{equation}
F(X) = F_{\rm pert}(X) + F_{\rm np}(X)\,.
\end{equation}
At large volume, one has \cite{Candelas:1990rm,Hosono:1994ax}
\begin{align}
\label{prepotcompactcy}  
F_{\rm pert}(X) &= i(X^0)^2\left(\frac{1}{3!}C_{ijk} T^i T^j T^k+\frac{\zeta(3)}{2}\chi(\text{CY}_3) \right),\\
F_{\rm np}(X)&=-i(X^0)^2\sum_{\vec\beta\in H_2(\rm{CY_3},\mathbb Z)} \alpha^{\vec \beta}_0\,\,\,{\rm Li}_3\left(e^{-\vec\beta \cdot\vec T}\right),
\end{align}
where $\chi(\text{CY}_3)$ is the Euler characteristic of the Calabi-Yau threefold, while $\alpha^{\vec \beta}_0$ are genus-zero Gopakumar-Vafa invariants \cite{Gopakumar:1998ii,Gopakumar:1998jq}.
The latter count the number of BPS configurations from the $D2$-branes wrapping a (genus-zero) curve in the class $\vec\beta\in H_2(\rm{CY_3})$.
That the terms above arise at string tree-level, i.e. at genus zero, can be seen after the gauge fixing $|X^0| \sim M_{\rm pl} \sim 1/g_s$ in string units, where $g_s$ is the string coupling. 
Notice that in $F_{pert}(X)$ we are omitting those linear and quadratic terms in $T^i$ that can be absorbed by a symplectic transformation \cite{deWit:1992wf,Harvey:1995fq}.

So far, we have discussed (non-)perturbative corrections in the string coupling constant to a two-derivative four dimensional $N=2$ supergravity action. It is known that the Wilsonian effective action of type IIA Calabi-Yau compacifications contains also a whole series of higher derivative corrections. 
They can be encoded into a holomorphic function
\begin{equation}
  \label{fullprephol}
\mathcal{F} = \sum_{g=0}^\infty \mathcal{F}_{g}(X)\,  \mathcal{W}^{2g}\,,
\end{equation}
where $\mathcal{W}^2$ is the square of the lowest component of the Weyl multiplet introducing higher derivative terms into the supergravity action \cite{LopesCardoso:1998tkj}.\footnote{Notice that in four dimensions a derivative expansion is not an expansion in $\alpha'$, since the moduli also contain $\alpha'$.} 
Due to the decoupling of the dilaton from the vector multiplet moduli space, there cannot be (non-)perturbative corrections to a given $\mathcal{F}_g$ appearing at order $g'>g$. Hence, $\mathcal{F}$ is for us the complete holomorphic prepotential, while the object $\mathcal{F}_0\equiv F$ corresponds to the two-derivative (non-)perturbative couplings discussed before. 
The corrections $\mathcal{F}_g$ are intimately related to topological string amplitudes at genus $g$. The precise relation is nevertheless subtle and we refer to \cite{Cardoso:2014kwa} for more details. What will be important for us in the following is that $\mathcal{F}_g$ can be calculated by counting BPS states as originally proposed in the seminal work of Gopakumar and Vafa  \cite{Gopakumar:1998ii,Gopakumar:1998ki,Gopakumar:1998jq}.

The holomorphic terms in \eqref{fullprephol} contribute to higher derivative interactions of the type $\mathbb{F}_g(T,\ov{T}) R_{+}^2 F_{+}^{2g-2}$, where $R_{+}$ and $F_{+}$ are respectively the self-dual parts of the Riemann tensor and of the graviphoton field strength. 
Up to an additive term independent of the K\"ahler moduli, the coupling splits into a harmonic piece and the so-called holomorphic anomaly
\begin{equation}
\mathbb{F}_g(T,\ov{T})={\rm Re}( \mathcal{F}_{g}(T) ) +  f_g^{anom}(T,\ov T)\, ,
\end{equation}
with $\partial_i\partial_{\bar {\jmath}}  f_g^{anom}(T,\ov T)\ne 0$.
Of particular interest for us is the term with $g=1$, which governs the coefficient of an $R^2$-correction \cite{Maldacena:1997de}, arising from an $R^4$-term in eleven dimensions \cite{Green:1997tv,Antoniadis:1997eg}. Indeed, such term a gives an estimate for the number of species
\begin{equation}
\label{Nsp as F_1}
\mathbb{F}_{1} \simeq N_{\rm sp}\,,
\end{equation}
as recently proposed in \cite{vandeHeisteeg:2022btw} and subsequently developed in
\cite{Cribiori:2022nke,vandeHeisteeg:2023ubh,vandeHeisteeg:2023uxj,Cribiori:2023ffn,Cribiori:2023sch}.

The moduli-dependent expression of the number of species thus obtained is valid over the entire vector multiplet moduli space. 
We propose and we will make concrete in examples that the quantity
\begin{equation}
g_E\simeq \frac{1}{\mathbb{F}_{1}} \simeq \frac{1}{N_{\rm sp}}
\end{equation}
naturally provides a small parameter in which to organize pertubatively the various quantum gravity theories emerging in the asymptotic regions of the moduli space. 
In other words, the coupling of the emergent theory, $g_E$, will serve as a compass in our analysis.

\subsubsection*{The resolved conifold}

A setup that will be highly illustrative for our purposes is the resolved conifold, which has been analysed in \cite{Gopakumar:1998ki} by developing a duality to Chern-Simons theory already pointed out in \cite{Gopakumar:1998ii}.
In this case, we have  only a single 2-cycle of topology $S^2$ so that the holomorphic prepotential takes a very simple form
\begin{align}\label{conifold g=0}
\mathcal{F}_{0} &=-\frac{1}{g_s^2}\left(\zeta(3)-\frac{\pi^2}{6}T-i\left(m+\frac{1}{4}\right)\pi T^2+\frac{T^3}{12}-{\rm Li}_3(e^{-T})\right)\\[0.3cm]
\label{conifold g=1}
\mathcal{F}_{1} &=\frac{T}{24}-\frac{1}{12}{\rm Li}_1(e^{-T})=\frac{T}{24}-\frac{1}{12}\sum_{m \geq 1} \frac1m e^{-mT}\,, 
\end{align}
where  $T$ is the complexified K\"ahler parameter of the $S^2$.
Comparing this with \eqref{prepotcompactcy}, one realizes that formally the self-intersection number of the $S^2$ is $C=1/2$.
For
$g>1$, one gets
\begin{equation}\label{conifold g>1}
\mathcal{F}_g=g_s^{2g-2}\left((-1)^g\chi_g\frac{2\zeta(2g-2)}{(2\pi)^{2g-2}}+\frac{\chi_g}{(2g-3)!}{\rm Li}_{3-2g}(e^{-T})\right)\,, 
\end{equation}
where $\chi_g$ is the Euler characteristic of the moduli space of all Riemann surfaces with genus $g$ and $h$ punctures. While the integer $m$ in \eqref{conifold g=1} is naively related to $h$ as $m=\frac{h}{2}-1$, it is actually not uniquely fixed.
Note that these higher genus $\mathcal{F}_g$ were derived in \cite{Gopakumar:1998ii} by computing a Schwinger integral with a single-wrapped $D2$-brane bound to an arbitrary number of $D0$-branes in the loop. This was the basic building block for a generalization to compact Calabi-Yau manifolds, where one eventually sums over all 1/2-BPS $D2$-branes wrapping 2-cycles in $H_2({\rm CY}_3,\mathbb Z)$.

\section{Decompactification limit}
\label{sec:decomp}

We begin by examining decompactification limits, the analysis of which will prove to be richer than expected.
First, we identify the towers of states becoming light in this limit. Then, we employ a Schwinger computation to show how the prepotential emerges by integrating these states out. Crucially, to recover the known expression, we will need to include all those entire towers with a typical mass scale not larger than the species scale.
Finally, we compare and contrast our methodology with the simplified EFT approach to the Emergence Proposal, where one usually integrates out only those states with masses up to the species scale.

\subsection*{Towers of light states}

For the sake of our analysis, it is sufficient to consider the standard example used to motivate the Emergence Proposal, namely a Calabi-Yau manifold, $Y$, with a single complexified K\"ahler modulus, $T=t+ib$. 
A similar analysis can be performed for the more generic case with an arbitrary number of K\"ahler moduli. 
The infinite distance limit we are interested in is $t\to\lambda t$ and $\lambda\to \infty$.\footnote{One can naturally think of $\lambda$ as the ratio between the final and the initial value of the vacuum expectation value $t_0$ along the trajectory over the moduli space. Besides, here and in the following, when writing $t$ we actually refer to $t_0$ and omit the subscript for convenience.}
This is accompanied by a rescaling of the string coupling,
$g_s\to\lambda^{3/2} g_s$, in such a way that the four-dimensional
dilaton, $\sigma \simeq M_{\rm pl}/M_s\simeq \mathcal{V}^{1/2}/g_s$,
remains constant.  Hence, we are dealing with strong (string) coupling limit.
Here and in the following, $M_s$ is the string mass while $\mathcal{V}$ is the volume of the Calabi-Yau threefold in units of the string length, which in this example is $\mathcal{V}\simeq t^3$.
Note that from the dual perspective of M-theory compactified on $Y\times S^1$ this limit is precisely the decompactification limit of the $S^1$, i.e. $R_{11}\to\lambda R_{11}$.
The question is which states have a typical mass scale not larger than the quantum gravity cut-off in this  decompactification limit.

The lightest degrees of freedom in the spectrum are particle-like and given by the tower of BPS bound states of $D0$-branes with mass 
\begin{equation}
m^{D0}_n=n(\Delta m)_{D0}\,,\qquad (\Delta m)_{D0} =\frac{M_{\rm pl}}{ {\cal V}^{\frac12}}\,.
\end{equation}
Since these are BPS states, we can trust their mass also for large
$g_s$, so that to this tower of states we associate the typical mass scale
\begin{equation}
{\mathfrak m}_{D0}= \frac{M_{\rm pl}}{{\cal V}^{\frac12}}\simeq \frac{M_{\rm pl}}{\lambda^{\frac 32}}\,.
\end{equation}
By looking at the $R^2$-correction in the spacetime effective action, we can read off the species scale. In the limit of interest, we have $\mathbb{F}_1 \simeq N_{\rm sp} \simeq t\simeq{\lambda}$ and thus \cite{vandeHeisteeg:2022btw}
\begin{equation}\label{species scale decomp}
\tilde\Lambda=\frac{M_{\rm pl}}{{\cal V}^{\frac16}}\simeq\frac{M_{\rm pl}}{\lambda^{\frac12}}\,.
\end{equation}
Alternatively, one can obtain the same result by recalling that for a decompactification from four to five dimensions the species scale is given by $\tilde \Lambda = M_{\rm KK}^{1/3}M_{\rm pl}^{2/3}$, and then by substituting $M_{\rm KK}$ with $(\Delta m)_{D0}$.
Due to the scaling of $g_s$, we can see that $\lambda\to\infty$ is a
strong coupling limit in the type IIA frame. However, for the emergent
theory we have in mind a perturbative expansion in  the small parameter $g_E\simeq 1/N_{\rm sp} \simeq \lambda^{-1}$.

Since $\tilde\Lambda$ is parametrically larger than $\mathfrak m_{D0}$, there is room for  additional degrees of freedom to become light.
In this respect, consider a $D2$-brane wrapped around a 2-cycle in
$H_2(Y,\mathbb Z)$. We get towers of particle-like states with typical
mass scale
\begin{equation}
\label{D2gap}
{\mathfrak m}_{D2}=\frac{M_{\rm pl}}{ {\cal V}^{\frac12}} t \simeq \frac{M_{\rm pl}}{\lambda^\frac12}\,,
\end{equation}
which is of the same order as the species scale. 
Since there will be only a few such states below the species scale, the latter will not change parametrically by their inclusion.
We note that $D0$- and $D2$-branes are the objects which are electrically charged under the graviphoton and the gauge fields residing in the $N=2$ vector multiplets, respectively.

However, this is not the end of the story, as there are more towers of states whose mass spacing is parametrically not larger than the species scale. 
First, we expect to have towers of Kaluza-Klein modes along the
Calabi-Yau threefold, whose mass spacing and typical mass scale is
\begin{equation}
{\mathfrak m}_{\rm KK}\simeq \frac{M_{\rm pl}}{ {\cal V}^{\frac16} \sigma}  \simeq \frac{M_{\rm pl}}{\lambda^\frac12 \sigma}\,.
\end{equation}
For constant $\sigma$, this is parametrically of the same order as the
species scale.

These are all the light particle-like towers that we found. However,
we can also have  BPS $NS5$-branes wrapped on a 4-cycle $\Sigma$ in $H_4(Y,\mathbb Z)$. 
They give rise to $h_{11}(Y)$ emergent strings in four-dimensions
\cite{Font:2019cxq} with tension
\begin{equation}
T_{\rm str}\sim \frac{M_s^2}{g_s^2} {\rm vol}(\Sigma)\sim M_{\rm pl}^2\frac{{\rm vol}(\Sigma)}{ {\cal V}}\sim \frac{M_{\rm pl}^2}{ \lambda}\,, 
\end{equation}
where $ {\rm vol}(\Sigma)$ denotes the volume of the four-cycle
$\Sigma$ in units of the string length. 
We do not know what the quantization of this effective string in four dimensions and at strong string coupling $g_s$ is, but the natural scale for its excitations is certainly
\begin{equation}
{\mathfrak m}_{\rm str}\simeq T_{\rm str}^{\frac 12} \simeq \frac{M_{\rm pl}}{\lambda^{\frac 12}} \,.
\end{equation}
In case the four-cycle is the $K3$ fiber of a $K3$ fibered Calabi-Yau threefold, the wrapped $NS5$-brane is dual to a heterotic string at weak string coupling (see section \ref{sec:emstringlim}). 
Via duality, its quantization leads indeed to this mass spacing for its string excitations. 
In the generic case, we expect that the spectrum will likely not be
equidistant, but the mass levels will fluctuate around this typical value.
We note that we arrive at the same scale by just taking the sixth
route of the $NS5$-brane tension 
\begin{equation}
{\mathfrak m}_{NS5}\simeq T_{NS5}^{\frac 16} \simeq \frac{ M_{\rm pl}\, g_s^{\frac 23}}{ {\cal V}^{\frac 12}}\simeq \frac{M_{\rm pl}}{ \lambda^{\frac 12} }\,.
\end{equation}
Hence, in contrast to what we will find in section \ref{sec:emstringlim} for an emergent string limit, in the decompactification limit one should better not speak of an emergent string but of an emergent $NS5$-brane.\footnote{In \cite{Blumenhagen:2023xmk} we will argue that the four-dimensional theory discussed here can be understood as a compactification of an emergent perturbative M-theory on a Calabi-Yau threefold.}
If this reasoning is correct, then the wrapped $NS5$-branes can also contribute BPS excitations with mass at the species scale.

One can then check that all other BPS modes, like wrapped $D4$-branes or the type IIA perturbative string states, are heavier than the species scale. Indeed, in terms of $g_E \simeq 1/N_{sp} \simeq 1/\lambda$ the typical mass scale in their towers is 
\begin{equation}
{\mathfrak m}_{D4}\simeq \frac{\tilde\Lambda}{ g_E}\,,\qquad
{\mathfrak m}_{F1}\simeq \frac{\tilde\Lambda}{ g^{1/2}_E}\,,
\end{equation}
and they are non-perturbative in the regime $g_E \to 0$.

To summarize, in this decompactification limit we propose to find a new perturbative quantum gravity theory,  whose fundamental degrees of freedom are given by (towers of) bound states of $D0$-branes, wrapped $D2$-branes and  wrapped $NS5$-branes, possibly carrying quantized Kaluza-Klein momentum along the Calabi-Yau threefold. 
Since the wrapped $NS5$-branes give rise to $h_{11}(Y)$ strings in four-dimensions, this is not just a theory of particles. 
From the dual perspective of M-theory on $Y\times S^1$, the corresponding light degrees of freedom are bound states of transversal (to $S^1$) $M2$-branes and $M5$-branes carrying Kaluza-Klein momentum along the $S^1$ and possibly along $Y$. 
This will be further elaborated upon in \cite{Blumenhagen:2023xmk}.

We do not know what this new perturbative quantum gravity theory really is and how to quantize it. 
In the following, due to its relation to M-theory, we will call it $M_{\rm CY}$. 
Following the strategy outlined in the introduction, for our proposal of a new quantum gravity theory emerging at each infinite distance limit to make sense, we should be able to obtain the prepotential terms $\mathcal{F}_{g}$ directly within $M_{\rm CY}$. 
Furthermore, the result should be organized in a perturbative expansion in the small parameter $g_E\simeq \lambda^{-1}$. 
If this turns out to be possible, then it would represent a non-trivial check for our claim and it will shed light on the  role that the Emergence Proposal plays in this context.

\subsection{The Schwinger Computation}

In the spirit of the Emergence Proposal, we will  now make an attempt to recover the prepotential terms $\mathcal{F}_g$ from a one-loop computation within $M_{\rm CY}$. 
The states to be integrated out have been identified in the previous section: they are bound states of $D0$-branes, wrapped $D2$-branes, excitations of wrapped $NS5$-branes, carrying Kaluza-Klein momentum along the Calabi-Yau threefold.
The problem is well-posed, but the tools to address it in general are lacking at present. 
However, for the  holomorphic quantities of interest there is a simplification, for we only have to consider BPS states running in the loop. 
Indeed, in the context of the topological string, the derivation of
amplitudes at genus $g$
from $D2$-$D0$  BPS bound states has been carried out (from the dual M-theory perspective) in the seminal work of Gopakumar and Vafa \cite{Gopakumar:1998ii,Gopakumar:1998jq}.
In this sense, the Gopakumar-Vafa idea is in fact an example of an emergence computation and it is closely related to what we are aiming at.

From \cite{Gopakumar:1998ii, Gopakumar:1998jq} we recall that the full genus expansion of the topological amplitudes can be resummed by Schwinger-type integrals as
\begin{equation}
\label{schwingergen}
\sum_g \mathcal{F}_g {W}^{2g}= \sum_{\vec \beta, r, n} \alpha_r^{\vec\beta} \int_0^\infty \frac{ds}{s} \frac{W^2}{\left(2i\sinh\left(\frac{sW}{2}\right)\right)^{2-2r}} e^{-sZ_{\vec{\beta},n}}.
\end{equation}
Here, $\alpha_r^{\vec{\beta}}$ are genus $r\geq 0$ Gopakumar-Vafa invariants, $s$ the Schwinger parameter, $W = \langle \sqrt{\mathcal{W}^2}\rangle$ the vacuum expectation value of the graviphoton field strength and
\begin{equation}
  \label{centralchargeall}
  Z_{\vec{\beta},n}={\frac{M_s}{g_s}}\left( \vec\beta\cdot\vec T +in\right)
\end{equation}
is the mass (equal to the central charge) of the BPS state with contributions from the wrapping class $\vec{\beta} \in H_2(Y,\mathbb{Z})$ and from $n$ units of Kaluza-Klein momentum along the M-theory circle. In type IIA language, $Z_{\vec{\beta},n}$ is the mass of $D2$-branes wrapping $\vec{\beta}$ and bound to $n$ $D0$-branes; they arise from dimensional reduction of wrapped $M2$-branes on the circle. 
A few comments on formula \eqref{schwingergen} are in order.  
First, the factor $\sinh(sW/2)$ encodes the effect of a non-vanishing background field $W$, namely the graviphoton background. 
Second, to each $\mathcal{F}_g$ on the left hand side contribute only terms on the right hand side such that $g\geq r$.
The index $r$ can be understood from the five-dimensional perspective as labelling a massive BPS state with SO$(4)$-spin content $I_1 \otimes \sum_r \alpha_r^{\vec \beta} I_r$, where $I_1 = 2[0] \oplus [1/2]$ and $I_r = (I_1)^r$.  The simplest configuration of bound states of $D2$-$D0$-branes correspond to $r=0$, in which case the $D2$-brane is wrapping an $S^2$ \cite{Gopakumar:1998ii}.

If we want to employ these ideas, we face the problem that in the original work \cite{Gopakumar:1998ii,Gopakumar:1998jq} there is no explicit mentioning of BPS bound states involving also wrapped $NS5$-branes and Kaluza-Klein modes along the Calabi-Yau.
However, our previous swampland-based analysis suggests that their contribution is of the same mass scale as the pure BPS $D2$-$D0$ bound states, and thus we would expect them to contribute as well. 
We propose that this issue can be resolved by noting that the wrapped $NS5$-branes and the Kaluza-Klein modes are uncharged under the vector multiplets, which in type IIA arise from the RR sector. As a consequence, these states do not contribute to the  central charge, which will still be given by the expression \eqref{centralchargeall}.
Besides, since the bound states remain BPS also when the $NS5$-branes and the Kaluza-Klein modes are included, their mass will coincide again with $Z_{\vec{\beta},n}$.
In short, considering more general bound states with also $NS5$-branes and Kaluza-Klein modes does not change the functional dependence of ${\cal F}_g$, but only  contributes to the number of BPS states, i.e.~the Gopakumar-Vafa invariants $ \alpha_r^{\vec\beta}$. 
Formally these are infinitely many states, so that a regularization has to be performed. 
However, like for string theory, a proper (so far unknown) quantization of the $NS5$-brane will presumably take care of it.
Since, to our knowledge, the Gopakumar-Vafa invariants have not yet been computed from first principles in type II string theory by literally counting BPS states on a Calabi-Yau, our swampland-based arguments suggest that these integers will also involve contributions from BPS states involving wrapped $NS5$-branes and potentially Kaluza-Klein modes along the Calabi-Yau.\footnote{Recently, impressive progress has been accomplished in computing Gopakumar-Vafa invariants at genus zero \cite{Carta:2021sms,Demirtas:2023als}, extending on \cite{Hosono:1993qy,Hosono:1994ax}. Since the employed methodology exploits mirror symmetry to extract the invariants from the periods of the mirror Calabi-Yau, we would call this an indirect computation, in contrast to actually counting $D2$-$D0$ BPS bound states. In view of this, we believe it to be possible that the Gopakumar-Vafa invariants determined in this indirect way include contributions from $NS5$-branes and Kaluza-Klein states as well. Additionally, a formula to extract directly the degeneracy of states (whose weighted sum enters the Gopakumar-Vafa invariants) on K3 fibered Calabi-Yau threefolds has been conjectured in \cite{Katz:2014uaa} (building on \cite{Katz:1999xq}) and tested recently in \cite{Cota:2020zse}. }

The above story is expected to be similar to the pattern of perturbative BPS states of the heterotic string on $K3\times T^2$.
In this case, the central charge of the supersymmetry algebra is given by the right moving Kaluza-Klein momentum $p_R$ along $T^2$ and the mass spectrum can be expressed as \cite{Harvey:1995fq}
\begin{equation}
  \label{bpshet}
\alpha' M^2={\frac12} p_R^2 + N_R-\frac12\,.
\end{equation}
Therefore, BPS states arise from right-moving ground states, $N_R=\frac12$, but can carry  unrestricted left-moving excitations $N_L$. 
This heterotic picture is precisely dual to what happens in the emergent string limit to be discussed in section \ref{sec:emstringlim}. 
For completeness, let us mention that the heterotic/IIA duality has been exploited in the seminal work \cite{Marino:1998pg} to compute topological amplitudes and to match the result with curve-counting formulae.

\subsection{Exact emergence on the resolved conifold}

We evaluate the Schwinger integrals for a single 2-cycle of genus zero, i.e.~a rigid $S^2$. For the (non-compact) resolved conifold this computation is already complete. There is also no compact 4-cycle so that there is no $NS5$-brane contribution. 
Notice that a 1/2-BPS $D2$-$D0$ bound state wrapping the $S^2$ cannot have Kaluza-Klein momentum along it.
This is the curved version of the familiar result that 1/2-BPS $D$-brane boundary states carry only momentum along transverse directions while winding in the longitudinal directions.\footnote{For the perturbative type II theories, this is also the case for Kaluza-Klein and winding modes of the fundamental string. Denoting with $p$ and $w$ the momentum and winding numbers respectively, only the configurations $(p,w)=(p,0)$ or $(p,w)=(0,w)$ are 1/2-BPS.}
Therefore, only bound states of $D0$-branes admit discrete Kaluza-Klein modes.

Hence, let us consider a bound state of a $D2$-brane wrapping the rigid 2-cycle and $n$ $D0$-branes, as in \cite{Gopakumar:1998ii}. Then, the mass spectrum is $ Z_{1,n} \equiv Z_n$ with
\begin{equation}\label{BPS charge}
Z_n =\frac{M_s}{\lambda^{3/2} g_s}\Big(\lambda t+i(b+n)\Big) = \frac{M_{\rm pl}}{\lambda^{3/2} \mathcal{V}^{1/2}} \Big( \lambda t+i(b+n)\Big)\,,
\end{equation}
and a pure  $D0$-tower is obtained by setting $t=b=0$ in the above expression.  
Since the rigid 2-cycle is topologically an $S^2$, the sum over $r$ in the Schwinger integral  \eqref{schwingergen} collapses to the value $r=0$ (genus zero) and, after expanding the denominator, we get\footnote{One may use $\frac{1}{\sinh^2 x } = - \frac{d}{dx} \coth x=\sum_{n=0}^\infty \frac{2^{2n}(2n-1)}{(2n)!} B_{2n} x^{2n-2}\,,\,0<|x|<\pi\,.$ }
\begin{equation}
 \mathcal{F}_{g} = - \frac{(2g-1)B_{2g}}{(2g)!} \sum_n \int_0^\infty ds\,s^{2g-3} e^{-s Z_n}\,,
\end{equation}
where the $B_{2g}$ denote Bernoulli numbers.  
As shown in \cite{Gopakumar:1998ii}, for $g>1$ the integral and the sum over all $n\in\mathbb Z$ can be carried out explicitly and turn out to be finite. 
For $g=0,1$, the objects we are interested in are at first sight divergent, so that one has to introduce an ultraviolet regulator, $\epsilon>0$, in the integral.
Therefore, we will be dealing with the quantities 
\begin{equation}
\mathcal{F}_1=-\frac{1}{12}\sum_{ n\in\mathbb{Z}}\int_{\epsilon}^{\infty}\frac{ds}{s} \,e^{-sZ_n}\,,\qquad \mathcal{F}_0= \sum_{n\in\mathbb{Z}}\int_{\epsilon}^{\infty}\frac{ds}{s^3} \,e^{-sZ_n},
\end{equation}
where we used $B_0=1$ and $B_2=1/6$.

To proceed, we first perform the integral over  $s$, then the sum over $n$. As for $\mathcal{F}_1$, we get
\begin{equation}
\label{F1 after int}
\mathcal{F}_1=\frac{1}{12}\sum_{n\in\mathbb{Z}} \Big(\gamma_E+\log(  \epsilon\, Z_n)+\mathcal{O}(\epsilon)\Big)\,,  
\end{equation}
where $\gamma_E$ is the Euler-Mascheroni constant.
As for $\mathcal{F}_0$, we get  
\begin{equation}
\label{F0 after int}
\mathcal{F}_0= \frac{1}{g_s^2}
\sum_{n\in\mathbb{Z}}\left(\frac{1}{2(\epsilon')^2}-\frac{g_s \,   Z_n}{\epsilon'}-\frac{ (g_s Z_n)^2 }{4} \Big( 2\log \left( \frac{\epsilon' Z_n}{g_s} \right) +2\gamma_E-3\Big)  +  \mathcal{O}(\epsilon')\right)\,,
\end{equation}
where we defined $\epsilon' = g_s \, \epsilon$ to display an overall multiplicative factor $g_s^{-2}$, matching with the resolved conifold \eqref{conifold g=0}. 
Next, we carry out the sum over the infinite number of $D0$-branes. 
Since it constitutes the main new quantitative result of the paper and it is central to our whole argument, we show the computation in some detail.

\subsubsection*{Evaluating $\mathcal{F}_1$}

Since it is simpler, we start from $\mathcal{F}_1$.
We substitute the central charge \eqref{BPS charge} into \eqref{F1 after int} and keep only the terms which are non-vanishing in the limit $\epsilon \to 0$, giving
\begin{equation}
\label{f1start}
\mathcal{F}_1^{D0,D2} = \frac{1}{12} \sum_{n \in \mathbb{Z}} \log \left( \frac{ T + in}{\mu}\right)\,
\end{equation}
with $\mu=e^{-\gamma_E}{\cal V}^{\frac 12}/(\epsilon M_{\rm pl})$. 
We omit the factors $\lambda$ in the calculation and re-introduce them
in the final result.
Then, the sum can be decomposed as
\begin{equation}
\label{f1evala}
\mathcal{F}_1^{D0,D2} = \frac{1}{12} \bigg( \log \left( T\right) + \sum_{n \geq 1} \log \left( 1+ \frac{ T^2}{n^2} \right)+ 2 \sum_{n \geq 1} \log(n) - \log (\mu) \sum_{n \in \mathbb{Z}} 1 \bigg)\,. 
\end{equation}
The second term on the right hand side can be evaluated with \eqref{log-sinh-id}, while the third and the fourth terms are divergent and need to be regularized. 
Next comes the crucial step, where instead of cutting these sums off at $n=N_{\rm sp}$ we keep all of the terms and exploit $\zeta$-function regularization.  As for the third term, we employ
\begin{equation}
\sum_{n \geq 1} \log(n) \equiv - \zeta ' (0) = \frac{1}{2} \log( 2\pi)\,,
\end{equation}
while for the fourth and last term we invoke
\begin{equation}
\sum_{n \geq 1} 1 \equiv \zeta (0)=-\frac{1}{2}\,,
\end{equation}
implying that
\begin{equation}
\label{nicezero}
\sum_{n\in\mathbb Z} 1\equiv 1+2\zeta(0)=0\ .
\end{equation}
Hence, thanks to this regularization scheme, the dependence of \eqref{f1evala} on the scale $\mu$ drops out, since the last term vanishes. Putting everything together, we obtain eventually
\begin{equation}
\begin{aligned}
\label{F1-D0-D2}
\mathcal{F}_1^{D0,D2} &= \frac{2\pi (\lambda t+ib)}{24} + \frac{1}{12} \log \Big(1-e^{-2\pi (\lambda t+ib)}\Big) \\[0.2cm]
&= \frac{2\pi }{24} \Big(\frac{t}{g_E}+ib\Big)- \frac{1}{12}  \sum_{m \geq 1} \frac{1}{m} e^{-2\pi m  \left(\frac{t}{g_E}+ib\right)} \,,
\end{aligned}
\end{equation}
where in the last step we recalled the series expansion of the logarithm and restored the scaling parameter $\lambda\simeq g_E^{-1}$.

We have just shown that by summing over the entire infinite tower of BPS states and treating the divergences via $\zeta$-function regularization one obtains the exact one-loop topological free energy of the resolved conifold \eqref{conifold g=1} (upon rescaling $T$ by $2 \pi$ appropriately to match conventions). 
In particular, this includes the term linear in $T$, which is usually obtained upon dimensional reduction \cite{Antoniadis:1997eg,Green:1997tv,Maldacena:1997de,Dedushenko:2014nya}.
Instead, we have been able to reconstruct it from the Schwinger integral using the light degrees of freedom of the emergent theory $M_{\rm CY}$.
To our knowledge, such a term has not been obtained in this manner before. 

To generalize the analysis to a compact Calabi-Yau threefold $Y$, one considers the limit $t_i\to\lambda t_i$ with
$\lambda\to\infty$ and $i=1,\ldots,h_{11}(Y)$. 
With the four-dimensional dilaton kept constant, this limit is dual to the decompactification limit to five-dimensions of M-theory on $Y\times S^1$.
The light degrees of freedom now include $D2$-$D0$ bound states wrapping all 2-cycles in the homology lattice $H_2(Y,\mathbb Z)$. 
Then, one must sum over them in the Schwinger integral so that the result will be sensitive to {\it all K\"ahler moduli} $t_i$. 
This means that the full topological amplitude ${\cal  F}_1$, including the linear terms, will emerge from a Schwinger one-loop computation.
Notice that the concrete evaluation might require yet another regularization for the sum over the linear terms in the K\"ahler moduli. 
As previously discussed, we expect also contributions from BPS states involving $NS5$-branes and Kaluza-Klein modes along $Y$. 
Determining their finite contributions requires a proper quantization of the wrapped $NS5$-branes, which has been so far out of reach.
Moreover, from the general Schwinger integral \eqref{schwingergen} one infers that there will be also similar contributions from $D2$-branes wrapping genus $g=r=1$ one curves in $Y$.

While this will certainly be an interesting future research direction, in order to proceed with the conifold example we need just to consider the contribution of a tower of pure $D0$-brane bound states. 
In this case, the central charge does not depend on $T$ so that we obtain\footnote{Note that compared to eq.\eqref{f1start} we have introduced by hand an overall relative minus sign in order to be in agreement with the extra $-\chi(Y)/2$ factor appearing for pure $D0$ bound states in \cite{Gopakumar:1998ii,Gopakumar:1998ki}. We expect this factor to arise from including BPS states with Kaluza-Klein modes along the $S^2$ followed by a proper regularization.} 
\begin{equation}
\mathcal{F}_1^{D0} = -\frac{1}{12} \sum_{n \ne 0} \log \left( \frac{in}{\mu}\right)\,.
\end{equation}
Notice that we excluded $n=0$, as some particle should really run in the loop. 
Proceeding with $\zeta$-function regularization and neglecting an irrelevant imaginary constant depending on the branch of the complex logarithm, we get
\begin{equation}
\mathcal{F}_1^{D0} = -\frac{1}{12} \log \left( 2\pi\mu\right)\,.
\end{equation}
Since $\mu$ contains a factor of ${\cal V}^{\frac 12}$,  it is tempting to speculate that this non-holomorphic logarithmic term is related to the holomorphic anomaly of $\mathcal{F}_1$. 
Furthermore, this time the result is not independent of the regulator $\epsilon$, arising due to the $\log(\mu)$ factor in $\mathcal{F}_1^{D0}$, but we can avoid this by subtracting this term from the final result.

\subsubsection*{Evaluating $\mathcal{F}_0$}

Next, we look at the prepotential $\mathcal{F}_0$. 
Using \eqref{nicezero}, one can see that at fixed (small) $\epsilon'$ the sum over the first two terms in the expansion \eqref{F0 after int} vanishes. 
The remaining terms can then be written explicitly as
\begin{equation}
\label{F0-D0-D2-1}
 \mathcal{F}_0 = -\frac{M_s^2}{2 g_s^2}\sum_{n \in \mathbb{Z}} (T + in)^2 \, \log \left( \frac{T + in}{\nu}\right)\,,
\end{equation} 
where $\nu=e^{-\gamma_E+{\frac32}}{\cal V}^{\frac12}/(\epsilon M_{\rm pl})$. 
We note that an expression of this kind arises also in the EFT approach to the Emergence Proposal, see e.g.~formula \eqref{gaugeloop}, where for instance the gauge couplings are computed via the one-loop corrections with the light tower of states running in the loop.
However, in that approach the tower is cut-off when the mass of the states exceeds the species scale.

To calculate $\mathcal{F}_0$, we first notice that the dependence on $\nu$ vanishes, due to \eqref{nicezero} and $\zeta(-2)=0$. 
Then, for a $D2$-$D0$ bound state we decompose the sum as 
\begin{equation}
\begin{aligned}
\label{F0-D0-D2-2}
\mathcal{F}_0^{D0,D2} =-{\frac{M_s^2}{2g_s^2}}\bigg[ &T^2 \log \left( T \right) + \sum_{n \geq 1} \left( T^2 - n^2\right) \log \left(1+ \frac{T^2}{n^2} \right) \\ 
&+2\sum_{n\geq 1}(T^2 -n^2)\log (n)-4\, T \sum_{n\geq 1}\,n\; {\rm arccot} \left( \frac{T}{n} \right)\bigg]\,,
 \end{aligned}
\end{equation}
where for the inverse trigonometric functions we use the principal domain.
To proceed, we first employ the identity \eqref{trilogid} and then regularize all infinite sums over $n$ via $\zeta$-function regularization.
Using in particular
\begin{equation}
2\sum_{n \geq 1} \left(  T^2 -n^2\right) \log (n) \equiv T^2 \log (2\pi) - \frac{\zeta(3)}{2 \pi^2}\,,
\end{equation}
we finally arrive at the simple expression
\begin{equation}
\label{prepotfin}
\mathcal{F}_0^{D0,D2}=\left({\frac{M_s}{2\pi g_s}} \right)^2 \left[ {\rm Li}_3(e^{-2\pi T})-\frac{(2\pi T)^3}{12}\right]\,,
\end{equation}
with $T=t/g_E+ib$.  
Note that after rescaling $T$ by $2\pi$ and setting $M_s=2\pi$ this is equal to the result \eqref{conifold g=0} for the resolved conifold up the non-physical linear and quadratic terms in $T$. 
In appendix \ref{appB:altder} we provide an alternative derivation employing the Hurwitz $\zeta$-function.

Finally, we consider a bound state of only $D0$-branes, which is given by
\begin{equation}
\mathcal{F}_0^{D0}=-\frac{M_s^2}{2 g_s^2}\sum_{n\ne 0}n^2\log\left(\frac{n}{\nu}\right)= -\left(\frac{M_s}{2\pi g_s}\right)^2 \zeta(3)\ .
\end{equation}
In contrast to the analogous contribution in $\mathcal{F}_1$, this term does not have any dependence on $\nu$ upon using once more $\zeta(-2)=0$. Hence, $\mathcal{F}_0^{D0}$ comes out holomorphic.

We have just shown that, similarly to $\mathcal{F}_1$, also the full prepotential $\mathcal{F}_0$ of the resolved conifold, including the polynomial terms, can be obtained by a loop computation in the emergent theory $M_{\rm CY}$. Once more, it has been essential to sum over the full towers of BPS states and appropriately regularize via $\zeta$-function the infinite sums. In addition, despite its simplicity this example has served as a proof of principle that exact emergence is possible via the (strong) Emergence Proposal when including all towers whose mass spacing is smaller or equal to the species scale. Had we neglected, say, the tower of $D2$-$D0$ bound states, which are heavier than the pure $D0$-brane tower but nevertheless below the species scale, we would not have reached exact emergence. 
Then, if these BPS states need to be included, it seems inconsistent to not include as well bound states involving $NS5$-branes.

Again, the generalization to compact Calabi-Yau manifolds would be very interesting.
As in the original work \cite{Gopakumar:1998ii,Gopakumar:1998jq}, it is tempting to just sum over all genus-zero contributions guaranteeing that one recovers the instanton sum of Gopakumar-Vafa.
However, whether also the cubic tree-level contribution to the prepotential
correctly emerges remains to be seen.\footnote{One might be sceptical that e.g.~the cubic terms for free toroidal orbifolds can appear in such a way. We thank the referee for pointing this out to us.}

For completeness, let us recall how these couplings usually arise from the perspective of the dual M-theory. 
The polynomial terms arise via dimensional reduction of the eleven-dimensional supergravity action, whereas the exponential terms come from $M2$-brane instantons wrapping the $S^1$ and a 2-cycle on $Y$. 
In the decompactification limit of an $S^1$ of infinite size, i.e.~in the five-dimensional theory, the contribution of these latter terms vanishes.

\subsection{Comparison with the Emergence Proposal}

In order to appreciate the difference between the calculation above and the simplified EFT approach usually taken in the Emergence Proposal, let us sketch how the latter would work for the case of $\mathcal{F}_0$. 
The logic would be to sum only over those states whose masses lie below the quantum gravity cut-off, i.e.~the species scale, $\tilde \Lambda$. 
In fact, the application of this method to the sums that we are working with has been already proposed in \cite{Castellano:2022bvr}: one is instructed to choose the scale $\nu=\tilde\Lambda$ and to explicitly cut-off the sum in \eqref{F0-D0-D2-2} at the level $n=N_{\rm  sp}={\cal V}^{1/3}$. 
This leads to the following expression for the prepotential
\begin{equation}
\begin{aligned}
\label{prepota}
\mathcal{F}_0= -\frac{M_s^2}{2g_s^2}\sum_{n=0}^{{\cal V}^{\frac13}} \Big[(T^2-n^2)\log\left({\frac{T^2+n^2}{ 2\, {\cal V}^{\frac23}}}\right) - 4n T {\rm arccot}\left({\frac{T}{n}}\right) \Big]\,,
\end{aligned}
\end{equation}
where a factor of $1/2$ has been introduced in the $\log$ for
convenience (in any case, we have not been keeping track of several
constant factors and normalizations). 
Proceeding then along with the logic same as in standard emergence computations, one considers the scaling $t\to \lambda t$,  ${\cal V}^{\frac13}\to \lambda {\cal V}^{\frac13}$, with $\lambda \to \infty$, and thus one is allowed to approximate the sum by an integral. 
Hence, one gets the leading order result
\begin{equation}
\begin{aligned}
 \mathcal{F}_0\approx -\frac{M_s^2\lambda^3}{18 g_s^2}\bigg[&3\pi
 (t^3-3t{\cal V}^{\frac 23}) +\Big(-6t^3+18t{\cal V}^{\frac23}\Big)  \arctan\left({\frac{t}{{\cal V}^{\frac13}}}\right) +\\[0.1cm]
  &\Big(-3t^2  {\cal V}^{\frac13} + {\cal V}\Big) \left(2+3\log 2 -3 \log\left(1+{\frac{t^2}{ {\cal V}^{\frac23}}}\right)\right)\bigg]\,.
\end{aligned}
\end{equation}
From this expression, it can be seen that only when ${\cal V}\simeq
t^3$, i.e.~for the one modulus case, one gets the simple leading order
expression
\begin{equation}
 \mathcal{F}_0\approx \frac{M_s^2}{ 18 g_s^2}(3\pi +4) (\lambda t)^3\,.
\end{equation}

This result agrees with the weak version of the Emergence Proposal, but compared to the exact emergence that we have been able to recover it features certain obvious shortcomings:
\begin{itemize}
\item The full $\mathcal{F}_0$ is supposed to be holomorphic, but the explicit cut-off for the sum introduces a non-holomorphic dependence on the moduli.
\item Even if a cubic polynomial part in $t$ arises, the prefactor is not correct and the higher-order corrections are certainly not only those known for the prepotential of type IIA Calabi-Yau compactification. Indeed, the computation of the exact discrete sum instead of the continuous integral would produce these exponential terms, but also several non-holomorphic corrections. 
\item The higher-order corrections arising when insisting on this approach would also depend on whether one takes the sum up to $N_{\rm sp}$ or to, say, $N_{\rm sp}+1$. In this sense, such a cut-off procedure is not well defined if one wants to extract more information than the qualitative leading cubic behavior.
\end{itemize}

In view of the very simple and yet successful $\zeta$-function regularization that we performed, we interpret all these issues as shortcomings of the too naive EFT  approach to the Emergence Proposal.
Instead, our computation shows that exact emergence is realized in this asymptotic decompactifcation limit if {\it all of the states in the towers with a typical mass scale $\mathfrak m$ up to the species scale $\tilde\Lambda$} are running in the one-loop diagram. 
This confirms quantitatively the intuition behind the Emergence Proposal of \cite{Palti:2019pca,Grimm:2018ohb}, for this limit.
Similarly to asymptotic string limits, such as the fundamental string at weak string coupling, the emergent perturbative quantum gravity theory that we have been probing does not need an explicit ultraviolet cut-off, but it provides finite answers for loop diagrams by its own. 
Recall that for the weakly coupled fundamental string, one also sums over all infinite string excitations and finiteness stems from modular invariance inducing a restriction of the integral over $\tau$ to the fundamental domain.

\section{Emergent String Limit}
\label{sec:emstringlim}

Given the concerns raised in \cite{Blumenhagen:2023yws} regarding the inclusion of string towers in the standard emergence calculations, one might expect that an emergent string limit shall be more complicated than the previous one. 
Instead, this limit is actually more familiar from the point of view of string dualities. Given that much of the following discussion is rather well-known, we can be brief and just touch upon the main points. 
There is an important difference between decompactification and emergent string limits. In the first case, as we have just seen, one can encounter multiple string-like objects which are actually emergent (wrapped) branes becoming asymptotic tensionless. In the second case, there is really only a single and fundamental string emerging. Mathematically, this is possible if the Calabi-Yau admits a certain $K3$ fibration.

In order to have a $K3$ fibration over a base $\mathbb{P}^1$, one needs to have at least two K\"ahler moduli, see e.g.~\cite{Aspinwall:1995vk}. 
A well-studied example is the Calabi-Yau $\mathbb{P}_{1,1,2,2,6}[12]$, which has Hodge numbers $(h_{21},h_{11})=(128,2)$ and for which the prepotential has an expansion
\cite{Candelas:1993dm,Hosono:1993qy}
\begin{equation}
 \label{F0-IIA}
 \mathcal{F}_0^{IIA} = T_2\, T_1^2 + \underbrace{(T_1^3 + \mathcal{O}(e^{-2\pi T_1}))}_{g^{(1)}(T_1)} + \mathcal{O}(e^{-2\pi T_2}) \,.
\end{equation}
Here, $t_2={\rm Re}(T_2)$ is measuring the size of the $\mathbb{P}^1$ which in the emergent string limit it is sent to infinity, i.e. $t_2 \rightarrow \lambda t_2$, with $\lambda \rightarrow \infty$. 
Again, we rescale $g_s \rightarrow \lambda^{1/2} g_s$ accordingly to keep the four-dimensional dilaton fixed and whenever referring to moduli we are implicitly looking at their vacuum expectation values.\footnote{In this limit, a natural way to define $\lambda$ is as the ratio between the final and initial value of $t_2$ along the trajectory over the moduli space that we are following.}

\subsubsection*{Towers of light states}

We start by identifying the species scale and the light towers of states in the strongly coupled region $\lambda \to \infty$.
Wrapping an $NS5$-brane on the $K3$ fiber yields a four-dimensional string of tension
\begin{equation}
T_{\rm str}\sim M_{\rm pl}^2 \frac{{\cal V}_{K3}}{ {\cal V}}\sim \frac{M_{\rm pl}^2}{ \lambda}\,,
\end{equation}
and whose excitations have a typical mass scale  
\begin{equation}
\mathfrak m_{\rm str}= {\frac{M_{\rm pl}}{\sqrt{t_2}}} \simeq \frac{M_{\rm pl}}{\sqrt{\lambda}}\,.
\end{equation}
Here, we denoted with $\mathcal{V}$ the volume of the Calabi-Yau, while $\mathcal{V}_{K3}$ is the volume of the $K3$ fiber. The latter is independent of the scaling of $t_2$.
In the presence of a string tower, the species scale is equal to the string mass scale itself, hence we identify\begin{equation}
 \tilde{\Lambda}  \simeq \mathfrak m_{\rm str}=  \frac{M_{\rm pl}}{\sqrt{\lambda}} \,.
\end{equation}

As opposed to the decompactification limit, in this case the sixth root of the $NS5$-brane tension is such that
\begin{equation}
\mathfrak m_{NS5}\sim T^{\frac 16}_{NS5}\sim  \frac{M_{\rm pl} g_s^{\frac 23}}{ {\cal V}^{\frac 12}}\sim  \frac{M_{\rm pl}}{ \lambda^{\frac16}}
\end{equation}
and hence gives a scale larger than $\mathfrak m_{\rm str} \simeq \tilde \Lambda$, providing another justification for talking about an emergent string limit in this case.
While there are no other states which are parametrically lighter than the species scale, there are a couple of  particle-like towers with a mass equal to it. 
First, due to the large size of $\mathbb{P}^1$, we have light Kaluza-Klein modes with a typical mass scale
\begin{equation}
\mathfrak m_{\rm KK} = \frac{M_s}{\sqrt{t_2}} \simeq \frac{M_{\rm pl}}{\sqrt{\lambda}} \,.
\end{equation}
The remaining light states are once again given by $D$-branes, namely $D0$-branes as well as $D4$-branes wrapped around $K3$ and $D2$-branes wrapped around the second homological 2-cycle, sitting inside the $K3$ fiber.
Their typical masses read
\begin{equation}\label{Dp part. string limit}
 \mathfrak m_{D0} = \frac{M_{\rm pl}}{\sqrt{t_2 \mathcal{V}_{K3}}}   \,, \qquad \mathfrak m_{D2} = \frac{M_{\rm pl}}{\sqrt{t_2 }} \,, \qquad \mathfrak m_{D4} = M_{\rm pl} \sqrt{\frac{\mathcal{V}_{K3}}{t_2}} \,
\end{equation}
and all of them scale as $M_{\rm pl}/\sqrt{\lambda}$.
Again, the wrapped $NS5$-branes and the Kaluza-Klein modes are uncharged under the three RR vector fields. 
Furthermore, note that a $D2$-brane wrapping the large base ${\mathbb P}^1$ is now too heavy to be included in the perturbative states.

\subsubsection*{Exact emergence of the prepotential}

Our claim of exact emergence, as defined in the Introduction, turns out to hold also in this new perturbative theory, where the full prepotential can be completely recovered again, even if not from a Schwinger integral exclusively. Let us stress that this notion of exact emergence has to be contrasted with an exact version of the Emergence Proposal in which everything is recovered just from one-loop integrals, as it happened in the previous case. As we will show, this is not possible in the present limit. Instead, the complete prepotential will arise via the correct quantization of the perturbative quantum gravity theory that emerges asymptotically. Indeed, we are presently in the convenient situation in which a known dual string description exists, namely the weakly coupled heterotic string, and we will exploit it to our advantage in order to recover the full prepotential. 
This is a special case of the $N=2$ heterotic/IIA duality in four dimensions \cite{Kachru:1995wm}.\footnote{Notice that one could realize exact emergence of the prepotential also in decompactification limits to M-theory, since the type IIA string coupling resides in a hypermultiplet. Our results instead point towards the realization of strong emergence in this limit, which is a much more non-trivial statement.}

The dual heterotic model is a compactifation on $K3\times T^2$.
Denoting the toroidal K\"ahler and complex structure moduli with $T$ and $U$ respectively, the dual model sits at the special locus $T=U$, where the Abelian gauge group $U(1)_U \times U(1)_T$ is enhanced to $SU(2) \times U(1)$.
The former type IIA abelian gauge fields can be identified with the heterotic gauge fields coming from the $T^2$ factor. 
Moreover, the particle-like states arising from wrapped $D0$-, $D2$- and $D4$-branes correspond to heterotic Kaluza-Klein and winding modes along the $T^2$. 
Kaluza-Klein modes on the $\mathbb{P}^1$ correspond to Kaluza-Klein modes along the $K3$ directions in the heterotic dual picture, hence they are uncharged under the gauge fields. 
The discussion can also be generalized to the case $T\neq U$, which requires a Calabi-Yau with $h_{11}=3$ and whose prepotential has been originally derived in \cite{deWit:1995dmj}.

With respect to the type IIA frame, a crucial difference is that the heterotic dilaton $S$ sits in a vector multiplet so that the prepotential can receive corrections at string one-loop and at the string non-perturbative level. Therefore, it takes the schematic form \cite{deWit:1995dmj,Kaplunovsky:1995tm}
\begin{equation}
\label{F0-het}
\mathcal{F}^{\rm het}_0 = \frac{1}{2} S T^2 + h^{(1)}(T) + h^{(\rm np)} (e^{-8\pi^2 S})\,.
\end{equation}
Here, the first term arises at string  tree-level, while the second is the dilaton-in\-de\-pen\-dent one-loop correction.
The function $h^{(1)}(T)$ is determined by its transformation properties under the exact quantum symmetry $SL(2,\mathbb{Z})_T$, acting on $T$ as in \eqref{modTtransf}, and by the singular structure at special points in moduli space. 
Modular invariance imposes that its second derivative reads
\begin{equation}
\del_T^2 h^{(1)}(T) = \frac{1}{4\pi^2} \log (j(iT)-j(i)) + \text{finite terms} \,,
\end{equation}
where $j(iT)$ is the modular invariant $j$-function (see appendix \ref{app_a} for more details).  
Microscopically, $ h^{(1)}(T)$ receives contributions from heterotic 1/2-BPS states which, as reviewed in the discussion around \eqref{bpshet},
also contain contributions from left-moving string excitations.
Via mapping the moduli according to 
\begin{equation}
T_{1} \rightarrow T \quad \text{and} \quad T_{2} \rightarrow 4\pi  S \,,
\end{equation}
it has been shown in \cite{Kachru:1995wm,deWit:1995dmj,Kaplunovsky:1995tm} that  the part $g^{(1)}(T_1)$ of the type IIA prepotential nicely matches the heterotic 1-loop correction $h^{(1)}(T)$. The other terms in \eqref{F0-het} are then trivially mapped to those in \eqref{F0-IIA}.

One might wonder what the role of the former type IIA 
Schwinger integral is in this setting. In the present case, when
expressed in terms of $M_s$, all light towers listed in \eqref{Dp part. string limit} are insensitive to the size of the large base, so that the sum over the Schwinger integrals will  depend only on the type IIA modulus $T_1$. 
The Kaluza-Klein modes on the large $\mathbb{P}_1$ do depend on $t_2$ but, just like the $NS5$-branes, they do not contribute to  the Schwinger-integral, as we have already explained in section \ref{sec:decomp}.
Hence, the (regularized) Schwinger integral only gives the  part $g^{(1)}(T_1)$ in the prepotential, which from the dual heterotic point of view is the string one-loop correction. This shows that the full prepotential is not emerging just from Schwinger integrals.

All this is supporting our claim that  in the asymptotic limit, we can consider the type IIA prepotential \eqref{F0-IIA} as a perturbative expansion in the small parameter $(\lambda \, t_2)^{-1}$, which is the inverse of the number of light species. Moreover, via duality we are also led to the conclusion that $g^{(1)}(T_1)$ does indeed receive contributions not only from $D2$-$D0$ bound states but also from excitations of $NS5$-$D2$-$D0$ bound states. The latter are dual to heterotic 1/2-BPS left-moving string excitations with Kaluza-Klein momentum and winding.

\section{Conclusion}
\label{sec:conclusion}

The Emergence Proposal is arguably among the less established swampland conjectures, but it may potentially have important consequences on our understanding of gravitational effective theories.  
When looking at it from a field theory perspective, one would usually let terms emerge in the low-energy effective action by integrating out (all) states whose masses are below the cut-off, which is upper bounded by the species scale.
While this strategy seems to give answers which are qualitatively correct, it nevertheless leads to several puzzles when quantitative tests are performed in concrete string setups \cite{Blumenhagen:2023yws}.

Motivated by this, in the present work we performed a quantitative analysis of the Emergence Proposal, aimed at understanding its precise place and formulation within string theory. 
Our results indicate that there exists an exact notion of emergence, which goes beyond the naive EFT approach usually employed in qualitative tests and extends emergence considerations to the full prepotential. With the tools available at present, while exact emergence seems to be observed in any asymptotic limit in moduli space, it is not always possible to recover all terms including the tree level ones from a one-loop integral, as one might expect from the original strong Emergence Proposal of \cite{Palti:2019pca,Grimm:2018ohb}.

Concretely, we focussed on the K\"ahler moduli space of type IIA string theory compactified on a Calabi-Yau threefold.
In this framework, we looked at both decompactification and emergent string limits. 
As for decompactification limits, we found that exact  emergence is indeed realized if one integrates out all states in towers with a typical mass scale  smaller than or equal to the species scale (here taken as the actual ultraviolet cut-off). 
This led us to the conclusion that considering only the lightest states, which would be just $D0$-branes in our setup, is certainly not sufficient, as it misses great part (and in fact almost all) of the moduli dependence.
Hence, one definitely needs to include wrapped $D2$-branes. In addition, our swampland arguments suggest that apart from them also wrapped $NS5$-branes and Kaluza-Klein modes should be included, as they have the same typical mass scale. 
This implies that the emergent (perturbative) quantum gravity theory in the decompactification limit is not just a theory of particles, for it includes also multiple four-dimensional strings arising from wrapped $NS5$-branes.

At the quantitative level, we showed that when summing over towers of states running in the loop diagrams, cutting off the sum at a certain level of the tower fails in reproducing the desired terms in the effective action: an integration over the full tower is needed.
It would be clearly important to collect more evidence to substantiate our proposal concerning the implicit inclusion of wrapped $NS5$-branes and Kaluza-Klein modes in the Gopakumar-Vafa invariants; we hope to come back to this problem in the future. 
However, even if this proposal will turn out to be incorrect, our computation of ${\cal F}_0$ and ${\cal F}_1$ for the conifold will still be valid and likewise our general conclusion on the  compatibility of the Emergence Proposal and exact emergence in decompactification limits.

As for emergent string limits, the situation appears to be simpler, since in the setup under investigation exact emergence is just another name for the well-known string-string duality, where once more one needs to consider infinite towers of states and no sharp cut-off is needed. However, realizing the strong Emergence Proposal in those limits appears to be technically out of reach at present.

More in general, we propose the following picture. In each of the asymptotic corners of the moduli space, assuming a fixed four-dimensional dilaton, a new  perturbation theory emerges which is organized by the inverse species number associated to that limit (recall that the species scale is a function of the moduli).
Then, one can either quantize this theory, a step which is typically out of reach at present, or hopefully understand it in terms of its dual, which is the strategy we followed in the present work. 
The nice aspect of the $N=2$ moduli space of type IIA string theory that we considered is that the exact solution for at least the holomorphic prepotential is known via non-renormalization theorems and mirror symmetry.
This solution holds even in the desert of the moduli space, where no towers of states are becoming light and all mass scales are close to the Planck scale.
Then, the various perturbative expansions with different light towers of states  are doomed  to reproduce that result in an expansion of the respective small parameter.

One can consider this moduli space in four dimensions as a well controlled raw-model for the quantum gravity theories in higher dimensions.
In particular, it will be  shown in \cite{Blumenhagen:2023xmk} that the decompactification limit to five dimensions studied in the present paper can be considered as a compactification on a Calabi-Yau of a perturbative description of M-theory in terms of transversal $M2$- and $M5$-branes carrying Kaluza-Klein momentum along the compact (large) eleventh directions.

Our analysis has implications on central ideas in the swampland program: it supports the existence of a unified portrait involving the distance conjecture, the \sout{(exact)} Emergence Proposal and the species scale, suggesting that any investigation focusing only on a subset of these ideas is necessarily incomplete. Furthermore, it provides also a novel perspective on the distance conjecture itself, which should be understood not really as a statement within effective field theory but rather about emergent (perturbative) theories of quantum gravity. To understand this point better, let us recall that in asymptotic limits, $\phi \to \infty$, there can really be three kind of towers of states with typical mass $\mathfrak m_{\rm tower}$:
\begin{itemize}
\item $\lim_{\phi\to \infty} \mathfrak m_{\rm tower}/\tilde \Lambda =\infty$, these states are decoupled in the emergent perturbative quantum gravity theory;
\item $\lim_{\phi\to \infty} \mathfrak m_{\rm tower}/\tilde \Lambda =0$, these states are part of the emergent perturbative quantum gravity theory;
\item $\lim_{\phi\to \infty} \mathfrak m_{\rm tower}/\tilde \Lambda =1$, these states are at the threshold of the ultraviolet cutoff and must be included in perturbative quantum gravity theory in order to have exact emergence. In particular, they do not decouple. 
\end{itemize}
Hence, one can argue that the swampland distance conjecture reflects a fundamental property of quantum gravity, namely that in asymptotic regions in field space there exists a perturbative description with a new emerging mass scale $\tilde\Lambda<M_{\rm pl}$ and whose consistent description requires to keep the full  towers of states with typical mass scale not larger than $\tilde\Lambda$.

Our results can be extended in various ways. One interesting extension would be to perform an orientifold projection on the setup and then study a moduli space with only four preserved supercharges, which is closer to phenomenological applications.
Alternatively, one can improve the quantitative aspects of our analysis. For example, one can trying to reconstruct precisely the triple intersection numbers (in $\mathcal{F}_0$) and the second Chern class (in $\mathcal{F}_1$) by summing  over all BPS bound states on the Calabi-Yau. Otherwise, one can also apply our methodology systematically to all possible asymptotic limits of the moduli space that we considered, which have been classified in \cite{Lee:2019wij}. 
Our analysis can also be repeated in compactifications letting a higher number of spacetime dimensions be non-compact; recent work in this respect has been performed in \cite{Etheredge:2023odp}. 
We hope to come back to these questions in the future.

\paragraph{Acknowledgments.}
We are grateful to  C.~Fierro Cota, T. Grimm, E.~Palti, I.~Valenzuela and T.~Weigand for discussions. The work of R.B.~and A.G.~is funded by the Deutsche Forschungsgemeinschaft (DFG, German Research Foundation) under Germany’s Excellence Strategy – EXC-2094 – 390783311.
The work of N.C.~is supported by the Alexander-von-Humboldt foundation.


\appendix

\section{Special functions}
\label{app_a}

In this appendix, we collect some useful functions and relations (see e.g.~\cite{NIST:DLMF}) that we employed in the main part of the work. 

\subsection*{$\zeta$-functions}

We recall the definition of the Riemann $\zeta$-function and its first derivative which, for ${\rm Re}(s)>1$, can be written as 
\begin{equation}\label{zeta definitions}
\zeta(s)=\sum_{n\geq1}{n^{-s}}\,, \qquad\zeta'(s)=-\sum_{n\geq 1}\frac{\log(n)}{n^s}\,.
\end{equation}
Outside ${\rm Re}(s)>1$ an analytic continuation is understood.
The trivial zeros of the $\zeta$-function are located at the negative integers, while for its derivative evaluated at these points we have
\begin{equation}\label{zeta deriv to zeta}
\zeta'(-2n)= (-1)^n\frac{(2n)!}{2(2\pi)^{2n}}\zeta(2n+1),\qquad 0 \neq n\in \mathbb{N}\,.
\end{equation}
Some particular values that are of importance in our calculations are
\begin{equation}\label{zeta values}
\zeta(0)=-\frac{1}{2}\,,\quad\zeta(2)=\frac{\pi^2}{6}, \quad\zeta(-2)=0,\quad  \zeta'(0)=-\frac{1}{2}\log(2\pi)\,,\quad \zeta'(-2)=-\frac{\zeta(3)}{4\pi^2}\,,
\end{equation}
where $\zeta(3)=1,2020569\ldots$ is Ap\'ery's constant. The Riemann $\zeta$-function is generalized by the Hurwitz $\zeta$-function, $\zeta(s,z)$. For  ${\rm Re}(s)>1$ and $z\neq 0, -1,-2,\ldots$, such function and its derivative with respect to $s$ are represented by the series
\begin{equation}
\zeta(s,z) = \sum_{n\geq 0}{(n+z)^{-s}} ,\qquad \zeta^\prime (s,z)= - \sum_{n\geq 0}(n+z)^{-s} \log(n+z)\,,
\end{equation}
otherwise one can perform an analytic continuation for $s\neq 1$. 

\subsection*{Polylogarithms}
We recall the definition of a polylogarithm 
\begin{equation}
{\rm Li}_s(z)=\sum_{n\geq 1}\frac{z^n}{n^s}\,,
\end{equation}
which is valid for any complex number $s$ and for $|z|<1$. 
The natural logarithm is recovered for the specific value $s=1$,
\begin{equation}
{\rm Li}_1(z)=-\log(1-z)\,,
\end{equation}
while for ${\rm Re}(s)>1$ the $\zeta$-function corresponds to $z=1$
\begin{equation}
{\rm Li}_s(1)=\zeta(s)\,.
\end{equation}

\subsection*{Useful identities}

Recalling that 
\begin{align}
\label{log-sinh}
\log \left( \frac{\sinh(\pi z)}{\pi z} \right) &=  \log \left[ \prod_{n=1}^{\infty} \left( 1 + \frac{z^2}{n^2} \right) \right]\,,
\end{align}
one can prove the identity
\begin{equation}
 \label{log-sinh-id}
\sum_{n\geq 1}\log\left(1+\frac{z^2}{n^2}\right)=\log\left(\frac{e^{\pi z}-e^{-\pi z}}{2\pi z}\right)=\pi z-\log{2\pi z}+\log(1-e^{-2\pi z})\,,
\end{equation}
which is used to obtain \eqref{F1-D0-D2}. 
Furthermore, integrating \eqref{log-sinh-id} twice and using
\begin{equation}
\int_0^A dz \, {\rm Li}_n(e^{2\pi z})=\frac{1}{2\pi}\left(-\zeta(n+1)+{\rm Li}_{n+1}(e^{2\pi A})\right)\,,
\end{equation}
one can derive the curious identity 
\begin{equation}
\begin{aligned}
\label{trilogid}
\sum_{n\geq
  1}\Big[\left(\frac{z^2-n^2}{2}\right)&\log\left(1+\frac{z^2}{n^2}\right)+
2zn \arctan\left(\frac{z}{n}\right)-{\frac 32} z^2\Big]=\\
 -&\frac{1}{4\pi^2} {\rm Li}_3(e^{-2\pi z})-\frac{\pi
   z}{12}+\frac{3}{4} z^2+\frac{\pi z^3}{6}-\frac{1}{2} z^2\log(2\pi z)+\frac{\zeta(3)}{4\pi^2}\,,
 \end{aligned}
\end{equation}
which is used to obtain \eqref{prepotfin}.
Another useful identity relates derivatives of the Hurwitz $\zeta$-function, polylogarithms and Bernoulli polynomials $B_n(x)$ as \cite{10.1145/258726.258736,MILLER1998201,Fucci:2011mf}
\begin{equation}
\label{idzetaprime}
\zeta'(-s,z) +(-1)^s \zeta'(-s,1-z) = \pi i \frac{B_{s+1}(z)}{s+1}+e^{-\pi i s/2} \frac{s!}{(2\pi)^s}{\rm Li}_{s+1}(e^{2\pi i z})\,.
\end{equation}
This is employed in appendix \ref{appB:altder} to provide an alternative derivation of the objects $\mathcal{F}_0$ and $\mathcal{F}_1$ for the resolved conifold.

\subsection*{Modular forms}

The modular group maps a complex modulus $T$ according to
\begin{equation}
\label{modTtransf}
iT \rightarrow \frac{iaT + b}{icT + d} \quad \text{with} \quad ad - bc  = 1 \,, \quad a,b,c,d \in \mathbb{Z} \,.
\end{equation}
A modular form $F_r (T)$ of weight $r$ is a holomorphic function from the upper half plane transforming as
\begin{equation}
F_r (T) \rightarrow (icT + d)^r F_r(T)
\end{equation}
under modular transformations. 

One example is the modular invariant $j$-function, defined for ${\rm Re}(T)>0$ as
\begin{equation}
j(iT) = \frac{E_4^3(iT)}{\eta^{24}(iT)} = \frac{E_6^2(iT)}{\eta^{24}(iT)} + 1728 \,.
\end{equation}
It is a combination of modular forms of weight 4 and 6, called Eisenstein series, which, for $q = e^{-2\pi T}$, are given by
\begin{equation}
E_4(iT) = 1 + 240 \sum_{n=1}^{\infty} \frac{n^3 q^n}{1-q^n} \,, \quad E_6(iT) = 1 - 504 \sum_{n=1}^{\infty} \frac{n^5 q^n}{1-q^n}
\end{equation}
and the Dedekind $\eta$-function
\begin{equation}
\eta(iT) = q^{\frac{1}{24}} \prod_{n=1}^{\infty} (1-q^n)\,,
\end{equation}
which is of weight 1/2, so that the $j$-function is of weight 0 and hence called the $j$-invariant. One of its properties that is particularly useful in our context is given by Borcherds' product formula \cite{Borcherds92}:
\begin{equation}\label{Borcherd prod}
j(iT)-j(iU)=p^{-1}\prod_{m>0,n\in\mathbb{Z}}(1-p^mq^n)^{c(mn)}\,,
\end{equation}
where $p=e^{-2\pi T}$ and $q=e^{-2\pi U}$ and the coefficients $c$ are given by the expansion
\begin{equation}
j(iT)-744=\sum_{n=-1}^\infty c(n)q^n=q^{-1}+196884q+21493760q^2\ldots\,.
\end{equation}
The infinite product converges only for $|p|,|q|<e^{-2\pi}$ and $p\neq q$, while the infinite series converges for $|q|<1$, but the relations may be extended on the entire complex plane via analytic continuation.
Taking the logarithm of \eqref{Borcherd prod} we obtain
\begin{equation}\label{Log(j's)}
\log(j(iT)-j(iU))=2\pi T+\sum_{m>0,n\in\mathbb{Z}}c(mn)\log(1-e^{-2\pi (mT+nU)})\,.
\end{equation}
While $T=U$ is a branch singularity of the above expression, one can already recognize the functional behaviors justifying the equivalence of \eqref{F0-IIA} and \eqref{F0-het}.

\section{Alternative derivation of $\mathcal{F}_0$ and  $\mathcal{F}_1$}
\label{appB:altder}

In this appendix, we provide an alternative derivation of $\mathcal{F}_0$ and $\mathcal{F}_1$ for the resolved conifold by means of Riemann and Hurwitz $\zeta$-function regularization. We start from $\mathcal{F}_0$, whose computation is more involved.

\subsubsection*{Evaluating $\mathcal{F}_0$}

We look at the contribution from $D2$-$D0$ bound states, which reads
\begin{equation}
-\frac{2g_s^2}{M_s^2}\mathcal{F}_0^{D0,D2} = \sum_{n\in \mathbb{Z}} (T+in)^2 \log\left(\frac{T+in}{\mu}\right)\,,
\end{equation}
and it can be decomposed as
\begin{equation}
-\frac{2g_s^2}{M_s^2}\mathcal{F}_0^{D0,D2}=-\sum_{n\in\mathbb{Z}}(T+in)^2\log{\mu}+\sum_{n\in\mathbb{Z}}(T+in)^2\log(T+in)\,. 
\end{equation}
One can check that the first term vanishes thanks to \eqref{nicezero} and $\zeta(-2)=0$, so that we have
\begin{equation}
\begin{aligned}
-\frac{2g_s^2}{M_s^2}\mathcal{F}_0^{D0,D2}&=-\log(i)\sum_{n\in \mathbb{Z}}(n-iT)^2 -\sum_{n\in \mathbb{Z}} (n-iT)^2\log(n-iT).
\end{aligned}
\end{equation}
Now, we proceed by regularizing the first sum with the Riemann $\zeta$-function, while the second with the Hurwitz $\zeta$-function. In this way, the coefficient of $\log(i)$ vanishes (thanks to \eqref{nicezero} and $\zeta(-2)=0$ again) and we arrive at
\begin{equation}
\begin{aligned}
-\frac{2g_s^2}{M_s^2}\mathcal{F}_0^{D0,D2}&=-\sum_{n\geq 0}(n-iT)^2\log(n-iT)-\sum_{n>0}(n+iT)^2\log(n+iT)\\
&\quad -\log(-1)\sum_{n>0}(n+iT)^2.
\end{aligned}
\end{equation}
We regularize the first and the second sum with the Hurwitz $\zeta$, while the third one with the Riemann $\zeta$. Using $\zeta(-2)=0$, $\zeta(-1)=-\frac{1}{12}$, $\zeta(0)=-\frac12$ and $\zeta'(s,z)\equiv-\sum_{n\geq 0}(z+n)^{-s}\log(z+n)$, we have
\begin{equation}
\begin{aligned}
\label{F1altstep1}
-\frac{2g_s^2}{M_s^2}\mathcal{F}_0^{D0,D2}&=\zeta'(-2,-iT)+\zeta'(-2,iT) +(iT)^2 \log(iT)\\
&\quad-\frac16 T(2k_1+1)\pi-\frac{T^2}{2}(2 k_2+1)i\pi,
\end{aligned}
\end{equation}
where we introduced $k_1,k_2\in \mathbb{Z}$ to keep track of the monodromies of the complex logarithm. Specializing \eqref{idzetaprime} for $s=-2$ and $z=iT$, one finds
\begin{equation}
\zeta'(-2,+iT) +\zeta'(-2,-iT) =i\frac{\pi}{3}B_3(iT) - \frac{1}{2\pi^2}{\rm Li}_3(e^{-2\pi T}) - (-iT)^2\log (-iT),
\end{equation}
with the Bernoulli polynomial $B_3(iT) = \frac{i}{2}T+\frac 32  T^2 -iT^3$. Inserting this into \eqref{F1altstep1}, we get
\begin{equation}
\begin{aligned}
-\frac{2g_s^2}{M_s^2}\mathcal{F}_0^{D0,D2}&=-\frac{\pi}{6}\left(2k_1+2\right)T-\frac i2\pi (2k_2+4k_3+2)T^2 \\
&\quad+\frac{\pi}{3}T^3 -\frac{1}{2\pi^2}{\rm Li}_3(e^{-2\pi T}),
\end{aligned}
\end{equation}
where $k_3\in \mathbb{Z}$ has been introduced to keep track of the branch of another complex logarithm, $\log(-1)=(2k_3+1)i\pi$. Observe that the integers $k_i$ may be tuned so that the non-physical terms of quadratic and linear order in $T$ vanish.

\subsubsection*{Evaluating $\mathcal{F}_1$}

The same logic may be applied to $\mathcal{F}_1^{D0,D2}$, namely
\begin{equation}    \mathcal{F}_1^{D0,D2}=\frac{1}{12}\sum_{n\in\mathbb{Z}}\log\left(\frac{T+in}{\mu}\right)=+\frac{1}{12}\log \left(\frac{i}{\mu}\right)\sum_{n\in\mathbb{Z}}1+\frac{1}{12}\sum_{n\in\mathbb{Z}}\log(n-iT)\,.
\end{equation}
The first term vanishes due to \eqref{nicezero}, so we get
\begin{equation}
    \begin{aligned}        12\mathcal{F}_1^{D0,D2}&=\sum_{n=1}^{\infty}\log(n+iT)+\sum_{n=0}^{\infty}\log(n-iT)+\log(-1)\sum_{n=1}^{\infty}1\\
    &=-\zeta'(0,iT)-\zeta'(0,-iT)-\log(+iT)+\frac{1}{2}(2k_1+1)\pi i\,,\quad k_1\in\mathbb{Z}\,.
    \end{aligned}
\end{equation}
Setting this time $s=0$ and $z=iT$ in \eqref{idzetaprime}, we get
\begin{equation}
\zeta'(0,iT)+\zeta'(0,-iT)=\pi i B_{1}(iT)+{\rm Li}_1(e^{-2\pi T})-\log(-iT)\,,
\end{equation}
where the relevant Bernoulli polynomial is now $B_1(iT)=iT-\frac{1}{2}$\,. Thus,
\begin{equation}
    \begin{aligned}
\mathcal{F}_1^{D0,D2}&=\frac{1}{12}\left(-{\rm Li}_1(e^{-2\pi T})+\pi T-\frac{\pi i}{2}+(2k_{2}+1)\pi i+\frac{1}{2}(2k_1+1)\pi i\right)\\
&=\frac{1}{12}\left(-{\rm Li}_1(e^{-2\pi T})+\frac{(2\pi T)}{2}+\frac{1}{2}(4k_2+2k_1+3)\pi i\right)\,,\quad k_1,k_2\in\mathbb{Z}\,,
    \end{aligned}
\end{equation}
which, after rescaling $T\rightarrow T/2\pi$, is in agreement with \eqref{conifold g=1},
up to a modulus-independent additive constant. Once again the integers $k_i$ were added to keep track of the complex logarithm branch ambiguities and are only affecting non-physical terms.

\newpage

\bibliography{references}  
\bibliographystyle{utphys}

\end{document}